\documentclass[floatfix,showpacs,amsmath,amssymb,letterpaper,groupaddresses,superscriptaddress]{article}
\usepackage{graphicx}
\usepackage{latexsym}
\usepackage{amsmath}
\usepackage{amssymb}
\usepackage{epsfig}
\usepackage{graphicx}
\usepackage{graphicx}
\usepackage{amssymb,amsmath,times}
\usepackage{hyperref}
\usepackage{color}

\def\a{\alpha}
\def\b{\beta}

\def\be{\begin{equation}}
\def\ee{\end{equation}}
\def\ba{\begin{eqnarray}}
\def\ea{\end{eqnarray}}
\def\la{\langle}
\def\ra{\rangle}
\def\a{\alpha}

\def\b{\beta}

\def\sq{\square}

\def\h{\hskip 1cm}

\def\lo{\longrightarrow}

\usepackage{epsfig}
\usepackage{graphicx}
\begin{document}
\begin{center}
{\Large \bf Completeness of classical $\phi^4$ theory on 2D lattices}\\
\vspace{1cm} Vahid Karimipour\footnote{email: vahid@sharif.edu}\h
Mohamamd Hossein Zarei\footnote{email:mhzarei@physics.sharif.edu}
\vspace{5mm}

\vspace{1cm} Department of Physics, Sharif University of Technology,\\
P.O. Box 11155-9161,\\ Tehran, Iran
\end{center}
\vskip 3cm

\begin{abstract}
We formulate a quantum formalism for the statistical mechanical
models of discretized field theories on lattices and then show
that the discrete version of $\phi^4$ theory on 2D square lattice
is complete in the sense that the partition function of any other
discretized scalar field theory on an arbitrary lattice with
arbitrary interactions can be realized as a special case of the
partition function of this model. To achieve this, we extend the
recently proposed quantum formalism for the Ising model \cite{quantum formalism}
and its completeness property \cite{completeness} to the continuous variable
case.
\end{abstract}

\section{Introduction:}
The understanding that a single partition function can describe
different phases of matter, is rather recent and indeed as late
as 1930's, there was not a consensus among physicists that a
partition function can give a sharp phase transition. The works of
Kramers, Wannier, Onsager \cite{onsager, kramers, kaufman} and others gradually
established beyond doubt that in the thermodynamic limit,
singular behavior and phase transition can arise from a single
partition function based on a single model Hamiltonian. For
example the Ising Hamiltonian can describe both the ordered phase
of a ferromagnet and the disordered phase of a paramagnet. Near
the point of second order phase transition, even the details of
the model Hamiltonian do not matter and only some general
properties
like symmetries are important. \\

Decades of works on statistical mechanics models, inspired by the
above general understanding, has revealed even further fruitful
relations between different models. Besides the well-known
duality relations between the low and high temperature phases of
the Ising model, one can also mention the so-called vertex
models, which reduce to other models in different limits.\\

One can ask if there are certain statistical mechanical models
which are complete, in the sense that their partition function
reduce to the partition function of any other model in a suitable
limit? If this turns out to be the case, then we can imagine a
very large space of coupling constants and one single
Hamiltonian, i.e. the Ising model with inhomogeneous couplings,
so that when we move through this space, we meet new phases and
new models which at present are thought to be completely
unrelated. This will then be another forward in the unification
program mentioned above.\\

It seems that the answer to the above question may be positive.
Recent results \cite{quantum formalism, completeness, delgado, u1, newj, statistic, nest, nest2, other, vidal} brought about by merging of ideas from
statistical mechanics and quantum information theory, give
positive clues in favor of the above idea. These investigations
\cite{quantum formalism, completeness} have been made possible by establishing a link between
statistical mechanics and quantum information on the one-hand and
the new paradigm of Measurement-based Quantum Computation (MQC)
and the universality of cluster states for
MQC on the other \cite{mqc4, dmqc1, dmqc2, dmqc3, dmqc4, dmqc5, dmqc6, dmqc7, dmqc8, dmqc9}.\\

In a series of recent works, it has been shown that the Ising
model on two dimensional square lattices with complex inhomogeneous
nearest-neighbor interactions, is complete in the sense that the
partition function of all other discrete models with general
$k$-body interactions on arbitrary lattices can be realized as
special cases of the partition function of Ising model on a
square lattice which is polynomially or exponentially larger than
the original lattice.  The starting point of these developments
was the observation in \cite{quantum formalism} that the
partition function of any given discrete model can be written as
a scalar product,

\begin{equation}\label{par1}
  Z_G (J)=\la \a |\Psi_G\ra,
\end{equation}
where $\la \a|$ is a product state encoding all the coupling
constants $J$ and $|\Psi_G\ra$ is an entangled {\it graph} state,
 defined on the vertices and edges of a graph, and encoding the geometry of the lattice. The core concept of completeness is the fact that the $2D$ cluster state is universal. More
concretely we know that the graph state $|\Phi_G\ra$
corresponding to a graph $G$ can be obtained from an appropriate
cluster state $|\Psi_\square\ra$ corresponding to a rectangular
lattice (denoted by $\square$), through a set of adaptive
single-qubit measurements ${\cal M}$. The measurements being
single-qubit can be formally written as $|\a_M\ra\la \a_M|$,
where $|\a_M\ra$ is a product state encoding the qubits the bases
and the results which have been measured. Thus the totality of
measurements ${\cal M}$ transforms the cluster state as follows
\begin{equation}\label{measure}
  |\Psi_\square\ra\lo |\a_M\ra\la a_M|\Psi_\square\ra=|\a_M\ra\otimes |\Psi_G\ra,
\end{equation}
which shows that after disregarding the states of the measured
qubits $|\a_M\ra$, what is left is an appropriate graph state
\begin{equation}\label{relation}
|\Psi_G\ra=\la \a_M|\Psi_\sq\ra.
\end{equation}
Combination of this relation with (\ref{par1}), leads to the
completeness result mentioned above. That is one writes
\begin{equation}\label{relation}
Z_G(J)\equiv \la \a |\Psi_G\ra=\la \a , \a_M|\Psi_\sq\ra,
\end{equation}
and notes that $\la \a, \a_M|$ now encodes a set of generally
inhomogeneous pattern of interactions on the cluster state.
Therefore one has
\begin{equation}\label{relation}
Z_G(J)\equiv Z_\sq(J,J').
\end{equation}

In this way it has been shown that the 2D Ising model with
complex inhomogeneous couplings is complete \cite{completeness}. \\

Like any completeness result, an interesting question is whether
there are other complete models. The situation is reminiscent of
results on NP completeness of certain problems in computer
science \cite{np}. In this direction it has been shown in
\cite{delgado, newj} that the four dimensional $Z_2$ lattice gauge
theory, with real couplings, is complete for producing any spin
model in any dimension. This result was extended in \cite{u1} to
show that the four dimensional $U(1)$ lattice gauge theory with
real couplings can produce, to arbitrary precision, a large
number of continuous models (those whose Hamiltonian allow a
finite Fourier series). Certainly there may be many other
complete models which can be converted to each other. Exploration
of the set of complete models certainly will add to our insight
and to our power in
connecting different models with each other.\\

In this paper we show that the discrete form of $\phi^4$ field
theory, on a two dimensional rectangular lattice is also complete
in the sense that the partition function of any continuous model
on any graph with any type of interaction can be obtained, to
arbitrary precision, from a $\phi^4$ model with in-homogeneous
complex couplings on an enlarged
2D lattice. \\

The structure of this paper is as follows: First we gather the
necessary ingredients for our analysis in section (\ref{1}), i.e.
elementary facts about Continuous Variable (CV) states, operators
and measurements. We then reconsider in a new language three
types of continuous variable stabilizer states, namely the CV Kitaev
states, the CV extended Kitaev states and the ordinary graph
states and the relations between them.  In section (\ref{3}) we
introduce the quantum formalism for scalar field models on
arbitrary graphs and investigate properties of these models,
properties which are made transparent by using the quantum
formalism and are otherwise not easy to unravel. Then in section
(\ref{5}), we show that the free field theory on two dimensional
rectangular lattice is complete for free theories in the sense
that from its partition function, every other free field theory
on any graph can be obtained as a special case. Finally we show
that the $\phi^4$ field theory on 2D rectangular lattice is
complete, in the sense that its partition function reduces to the
partition function of any interacting model on any graph. As in
the Ising case, the price that one pays is that the coupling
constants of the
complete model should be inhomgenuous and complex.\\

\section{Preliminaries:}\label{1}
In this section we collect the preliminary materials necessary
for generalization of the quantum formalism to the Continuous
Variable (CV) case \cite{cv1, cv2, cv4, cv5}. First we review the definition of
Heisenberg-Weyl group, and the way a unitary operator can be
performed on CV state (a qumode) by measurements of an
appropriate graph state. We end this section with a note on
decomposition of CV unitary operators.

\subsection{The Heisenberg-Weyl Group}

The definition of CV stabilizer states starts with generalization
of the Pauli group to the continuous setting \cite{cvstab1, cvstab2, cvstab3, dstab, mqc5}. For one qumode, (a
term which replaces qubit in the continuous setting) the resulting
group is called Heisenberg-Weyl group $W$, whose algebra of
generators is spanned by the coordinate and momentum operators
satisfying $[\widehat{Q},\widehat{P}]=iI$. Thus modulo $U(1)$
phases, the group $W$ is the group of unitary operators of the
form $w(t,s)=e^{it\widehat{Q}+is\widehat{P}}$. Since the two
unitary operators
\begin{equation}\label{eq4}
\hat{X}(s)=e^{-is\widehat{P}}~~~~~~~and~~~~~~~\hat{Z}(t)=e^{it\widehat{Q}},
\end{equation}
have the simple relation
\begin{equation}
\hat{Z}(t)\hat{X}(s)=e^{ist}\hat{X}(s)\hat{Z}(t),
\end{equation}
multiplication of any two elements of $W$ can be recast in the
form $X(t)Z(s)$ modulo a phase. The Heisenbergy-Weyl group can be
represented on the Hilbert space of one particle, spanned by the
basis states $|y\ra_q$ (eigenstates of $\widehat{Q})$ or $|y\ra_p$
(eigenstates of $\widehat{P}$). On these states, the operators $X$
and $Z$ act as follows:
\begin{eqnarray}\label{pauli}
&&\hat{Z}(t)|y\ra_q = e^{ity}|y\ra_q, \h \hat{X}(s)|y\ra_p =
e^{-isy}|y\ra_p,\cr &&\hat{Z}(t)|y\ra_p = |y+t\ra_p, \h
\hat{X}(s)|y\ra_q = |y+s\ra_q.
\end{eqnarray}
\textbf{Remark:} Hereafter we denote the states $|y\ra_q$ simply
as $|y\ra$ and use $|y\ra_p$ for eigenstates of $\widehat{P}$ as
above. \\
The CV Hadamard operator is a unitary non-Hermitian operator
defined as
\begin{equation}\label{Had1}
  \hat{H}:=\int |y\ra_p\la y|_q dy\equiv \frac{1}{\sqrt{2\pi}}\int e^{ixy}|x\ra\la y|
  dxdy,
\end{equation}
from which we obtain
\begin{equation}\label{Had2}
  \hat{H}\widehat{Q}\hat{H}^{-1}=\widehat{P}, \h \hat{H}^{-1}\widehat{P}\hat{H}=\widehat{Q}.
\end{equation}
Moreover from (\ref{Had1}) we find
\begin{equation}\label{H}
  \hat{H}^2:=\int dx |-x\ra\la x|,
\end{equation}
which leads to $\hat{H}^4=I$. Therefore the square Hadamard
operator acts as the parity operator. This means that
\begin{equation}\label{Had3}
  \hat{H}^2\widehat{Q}\hat{H}^{-2}=-\widehat{Q}, \h \hat{H}^{-2}\widehat{P}\hat{H}^2=-\widehat{P}.
\end{equation}
The $n-$ mode Heisenberg-Weyl group is the tensor product of $n$
copies of $W$, i.e. $W_n:=W^{\otimes n}$ and all the above
properties are naturally and straightforwardly extended to $n$
modes. Of particular interest is the continuous variable
$\hat{CZ}$ operator which is defined as
$\hat{CZ}|x,y\ra=e^{ixy}|x,y\ra$, with the operator expression
\begin{equation}\label{}
  \hat{CZ}:=e^{i\widehat{Q}\otimes \widehat{Q}},
\end{equation}
and satisfying the relation
\begin{equation}\label{}
  (\hat{X}(t)\otimes I)\hat{CZ}=\hat{CZ}(\hat{X}(t)\otimes \hat{Z}(t)).
\end{equation}
We will also need a more general operator $\hat{CZ}$, namely
$\hat{CZ}(s)=e^{is\widehat{Q}\otimes \widehat{Q}}$ which has the
following relation
\begin{equation}\label{eqcz}
  (\hat{X}(t)\otimes I)\hat{CZ}(s)=\hat{CZ}(s)(\hat{X}(t)\otimes \hat{Z}(st)).
\end{equation}

\subsubsection{Single mode unitary operators induced by measurements}\label{sec0}
In this subsection, we review how a CV unitary operator can be
induced on a mode by measuring a suitable graph state. Here we
restrict ourselves to operators diagonal in the coordinate basis,
since this is the only type of operators which we encounter in
our analysis. We also use the basic result of MQC that certain
states are complete, in the sense that by a suitable sequence of
adaptive single-site
measurements on them, any other state can be reached.  \\

\begin{figure}
  \centering
  \epsfig{file=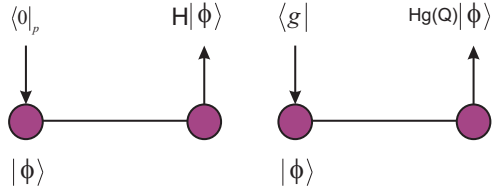,width=6.5cm}
  \caption{(Color online) Projection of the left mode on the zero momentum state is equivalent to the action of Hadamard operator on the right mode. Projection on the state $\la g|$ is equivalent to
  the action of the operator $\hat{H}g(\hat{Q})$ on the right mode. A downward arrow means projection on a state, an upward arrow means the resulting state.}
  \label{fig1}
\end{figure}

Consider a very simple two-mode graph state as shown in figure
(\ref{fig1}). The first mode is in an arbitrary state $|\phi\ra=\int dx
\phi(x) |x\ra$, the second mode is in the state
$|0\ra_p:=\frac{1}{\sqrt{2\pi}}\int dy |y\ra$, the two modes have
been joined by a CZ operator (shown by a line in figure(\ref{fig1})) and so
the two mode state is
\begin{equation}\label{}
  |\Psi\ra_{12}=(CZ)_{12}|\phi\ra_1|0_p\ra_2=\int dx dy e^{ixy}\phi(x) |x,y\ra_{1,2},
\end{equation}
where the indices 1 and 2 refer to the modes from left to right
in figure (\ref{fig1}). Now we project the first mode on the zero momentum
state $\la 0|_p$ \cite{mqc5, mqc, mqc2, mqc3, mqc6, mqc7, mqc8}. The state of the second mode will be
\begin{equation}\label{}
  |\phi'\ra_2=_1\la 0_p|\Psi\ra_{12}=\frac{1}{\sqrt{2\pi}}\int dy e^{ixy}\phi(x)
  |y\ra_{2}=H|\phi\ra.
\end{equation}
Thus projection of the first mode onto a zero momentum state is
equivalent to the action of the Hadamard operator on the state
$|\phi\ra$ and putting it on the second mode. This is shown in
figure (\ref{fig1}), where projection is depicted by a downward
arrow and
the result is depicted by an upward arrow.\\

Suppose now that we project mode 1 onto the state $\la
g|:=\frac{1}{\sqrt{2\pi}}\int dy g(y)\la y|$. If we note that
\begin{equation}\label{}
  \la g|=\frac{1}{\sqrt{2\pi}}\int dy g(y)\la y|=\la 0_p|g(\hat{Q}),
\end{equation}
and note that $g(\hat{Q}_1)$ commutes with $(CZ)_{12}$, we find
that projecting the first mode on the state $\la g|$ is
equivalent to the action of the operator $Hg(\hat{Q})$ on the
state $|\phi\ra$ and putting it on the second mode. This is shown
in figure (\ref{fig1}).
\begin{figure}
  \centering
  \epsfig{file=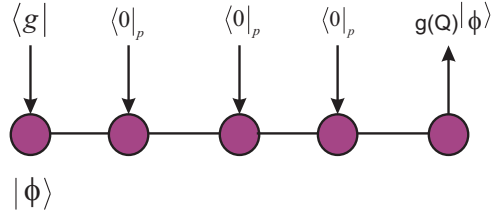,width=6.5cm}
  \caption{(Color online){The measurement pattern which enacts the operator $g(\hat{Q})$ on the right-most mode.}}
  \label{fig2}
\end{figure}
We can write this symbolically as
\begin{equation}\label{eqp0}
  P_0\lo H, \h P_g\lo Hg(\hat{Q}),
\end{equation}
where in the left hand side we show the projections and in the
right hand side we show the resulting action on the state.
 In order to enact the operator
$g(\hat{Q})$, i.e. remove $H$ from $Hg(\hat{Q})$, we need to
enact the operator $H$ three times. Thus using the symbols in
equation (\ref{eqp0}), we have
\begin{equation}\label{}
  P_0P_0P_0P_g\lo H^3 (Hg(\hat{Q}))=g(\hat{Q}),
\end{equation}
which is shown in figure (\ref{fig2}).\\

We are now in a position to state a basic theorem \cite{Braunstein, Lioyd, sornborger, Loock} in this
section. \\

\textbf{Thereom:} Let $V(\hat{Q})$ be a polynomial of $\hat{Q}$
with real coefficients and $|\phi\ra$ be an arbitrary state of an
appropriate chain of a cluster state. Then by projecting the
modes of this cluster state on the following three types of
states,
\begin{equation}\label{eqquad}
\la \beta_1(t)|:=\int dy e^{-ity}\la y|,\h \la \beta_2(t)|:=\int
dy e^{-ity^2}\la y|, \h \la \beta_4(t)|:=\int dy e^{-ity^4}\la y|,
\end{equation}
we can enact any operator of the form $e^{-iV(\hat{Q})}$ on the
state $|\phi\ra$, to any
desired precision. The state will appear on the un-projected modes of the chain. \\

\textbf{Proof:} First we note that enacting the operator $H$
comes for free by projecting on the zero momentum state $\la
0|_p$. Second we use the operator identity
\begin{equation}\label{}
  e^{tA} e^{tB} e^{-tA} e^{-tB}\approx e^{t^2[A,B]+o(t^2)}.
\end{equation}
From projection on the states $\la \b_i(t)|$ we find that the
operators $e^{-it\hat{Q}}$,  $e^{-it\hat{Q}^2}$ and
$e^{-it\hat{Q}^4}$ can be obtained. In view of the existence of
the Hadamard operator, the algebra of anti-Hermitian operators in
the exponential is generated by the set $\{iQ,iQ^2,iQ^4, iP, iP^2,
iP^4\}$. It is now easy to see that this algebra contains all
monomials of $Q$. To show this we first note that
\begin{equation}\label{Q3}
  [iP,iQ^4]\sim iQ^3,
\end{equation}
where $\sim$ means that we have ignored numerical factors. We
then note that
\begin{equation}\label{}
  [iP,[iP^2,iQ^4]]\sim i(PQ^2+Q^2P).
\end{equation}
 The
latter operator now acts as a raising operator for powers of
monomials, since
\begin{equation}\label{}
  [i(PQ^2+Q^2P),iQ^n]]\sim iQ^{n+1},
\end{equation}
which completes the proof. In this way we can generate any
unitary operator of the form $e^{-iV(Q)}$, where $V$ is a real
polynomial of $Q.$\\

\section{Three Classes of Continuous Variable States}\label{2}

Let $G=(V,E)$ be a graph, where $V$ and $E$ respectively denote
the set of vertices and edges. The graph is supposed to admit an
orientation. In other words, $G$ is the triangulation of an
orientable manifold.  This means that all the simplexes of $G$
inherit the orientation of the original manifold in a consistent
way. The number of vertices and edges are respectively given by
$|V|$ and $|E|$.\\

In this section we define three closely related continuous
variable states pertaining to a given graph $G$, which we call the
Kitaev state $|K_G\ra$ \cite{kitaev}, the extended Kitaev state
$|\overline{K}_G\ra$ and the graph state $|\Psi_G\ra$ \cite{graphstate, graphstate2}. We will
then determine the mutual relationships of these states, which
will play an important role in our proof of completeness. These
are the generalizations of known states in the qubit case, where
they have been possibly named differently in other works. For
example in the qubit case, extended Kitaev states have been called
pseudo graph states \cite{completeness}, however in view of their explicit
construction and stabilizers, we think that the name Kitaev or
extended Kitaev states are more appropriate for them.\\

The crucial difference between the continuous and qubit case is
the fact that the operators $\hat{X}(t)$, $\hat{Z}(t)$, $CZ$ and
$\hat{H}$ are not equal to their inverses. Therefore a consistent
and unambiguous description of these states on a graph, requires
that the graphs be decorated with weights and/or orientations. We
emphasize the difference between orientation, which is a $Z_2$
variable and weight which is a real variable. We will meet the
necessity of each as we go along in our definitions.

\subsection{Kitaev States}

Consider an oriented graph $G=(V,E,\sigma)$, where $\sigma$ means
that arbitrary orientations have been assigned to the edges. Any
collection of arbitrary orientations on the edges is called a
decoration of the graph. We assume that modes live only on the
edges $E$ and there are no modes on the vertices $V$ of this
graph. The CV Kitaev state is then defined
\begin{equation}\label{eqG'}
|K_G\rangle=\int d\phi_{1}
d\phi_{2}...d\phi_{N}\bigotimes_{e_{ij}}|\phi_i-\phi_j\rangle ,
\end{equation}
where $e_{ij}$ is the edge which goes from the vertex $i$ to the
vertex $j$. \\

It is easily verified that this state is stabilized by the
following set of operators: for each vertex $i\in V$, we have
\begin{equation}\label{eqA}
  A_v(t):= \prod_{e\in E_v} X_{e}^{\pm}(t),
\end{equation}
where $E_v$ denotes the set of edges incident on the vertex $v$
and the $-$ and $+$ signs are used for edges going into and out of
a vertex respectively. The reason that $A_v$ stabilizes the state
(\ref{eqG'}) is that it simply shifts the variable $\phi_v$ which will be
neutralized under the integration. Also for each face of the
graph, we have
\begin{equation}\label{eqB}
  B_f(s):= \prod_{e\in \partial f} Z_{e}^{\pm}(s),
\end{equation}
where $\partial f$ denotes the set of edges in the boundary of
$f$ and the $+$ and $-$ signs are used respectively when the
orientation of a link is equal or opposite to us when we traverse
a face in the counter-clockwise sense. Note that traversing all
the plaquettes in this sense is meaningful for an orientable
traingulation. Here also the effect of $B_p(s)$ on the state
inside the integral is to multiply it by a unit factor since the
phases acquired by all the edges add up to zero
for a closed loop. \\

As an example, we have for the graph ${\cal G}$ in figure (\ref{fig4}),
\begin{equation}\label{examk}
  |K_{\cal G}\ra=\int Dx |\phi_1-\phi_2, \phi_2-\phi_3, \phi_4-\phi_3, \phi_4-\phi_1, \phi_4-\phi_2\ra_{a,b,c,d,e},
\end{equation}
where the subscripts $a$ to $e$, determine the position of modes
in the state. The stabilizers of this state are then given by (\ref{eqA},\ref{eqB})
as follows:
\begin{figure}
  \centering
  \epsfig{file=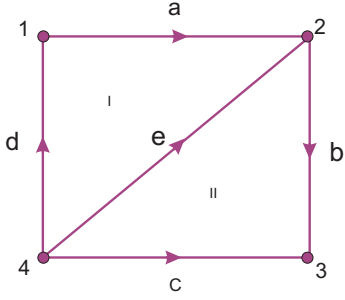,width=4.5cm}
  \caption{(Color online) The graph for which the Kitaev state is given in (\ref{examk}).}
  \label{fig4}
\end{figure}
\begin{equation}\label{kit1V}
  A_1:= X_d^{-1}X_a, \ \ \ A_2:= X_a^{-1}X_e^{-1}X_b,\ \ \  A_3:= X_b^{-1}X_c^{-1},\ \ \  A_4:=X_cX_dX_e
\end{equation}
and
\begin{equation}\label{kit1P}
  B_{I}:= Z_eZ_a^{-1}Z_d^{-1}, \h B_{II}:= Z_cZ_b^{-1}Z_e^{-1}.
\end{equation}

Finally we note that all Kitaev states on a given graph,
corresponding to different decorations are related to each other
by local unitary actions. In fact if we switch the arbitrary
orientation on a link $e_{ij}$, it means that the term $ |\cdots
\phi_i-\phi_j,\cdots \ra $ in (\ref{eqG'}) changes to $|\cdots
\phi_j-\phi_i,\cdots \ra$ where the remaining parts of the state
will remain intact. In view of (\ref{H}), this switching is achieved by a
local action of the square Hadamard operator $H^2$ on the edge
$e$. We can thus write
\begin{equation}\label{}
  |K_{G^{\sigma'}}\ra=\left(\bigotimes_{e: \sigma(e)\ne \sigma'(e)} H^{2}_e \right)
  |K_{G^{\sigma}}\ra.
\end{equation}
Therefore all the Kitaev states with different decorations belong
to the same class of states modulo local actions of square
Hadamard operations. \\

In view of the shift invariance $\phi_i\lo \phi_i+\eta$, the
Kitaev state has a hidden multiplicative factor which is in fact
infinite. This symmetry can be removed by fixing a gauge (in the
discrete case this is a finite factor which causes no problem).
Therefore we will define the gauge-fixed Kitaev state, denoted by
$|K^0_G\ra$ as follows:
\begin{equation}\label{eqG2'}
|K^0_G\rangle=\int d\phi_{1}
d\phi_{2}...d\phi_{N}\delta(\phi_1+\phi_2+\cdots \phi_N)
\bigotimes_{e_{ij}}|\phi_i-\phi_j\rangle.
\end{equation}
This gauge-fixed Kitaev state has still the same set of
stabilizers. The same reasoning for the Kitaev state $|K_G\ra$
also works here.
\subsection{Extended Kitaev States}
\begin{figure}
  \centering
  \epsfig{file=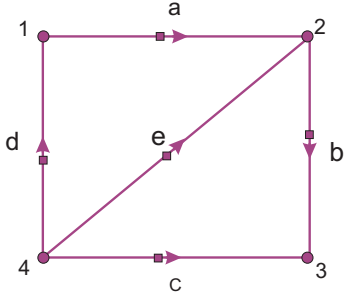,width=4.5cm}
  \caption{(Color online) The graph $G_0$ for which the extended Kitaev state is as in (\ref{eqek}).}
  \label{fig5}
\end{figure}

Let $G=(V,E,\sigma)$ be defined as in previous subsection. Now in
addition to the modes on the edges, there are also modes on
vertices, as in figure (\ref{fig5}).  On such a graph, there are thus two
different sets of vertices, which we denote by $V$ (the ones on
the nodes) and $V_E$ (the ones on the edges). Then the state
$|\overline{K}_G\ra$ is given by
\begin{equation}\label{eqeG}
|\overline{K}_G\rangle=\int d\phi_{1}
d\phi_{2}...d\phi_{N}\bigotimes_{e_{ij}\in E
}|\phi_{i}-\phi_j\rangle \bigotimes_{i\in V}|\phi_{i}\rangle,
\end{equation}
where $e_{ij}$ is the edge which goes from vertex $i$ to $j$.
 As
an example for the graph in figure (\ref{fig5}), we have
\begin{equation}\label{eqek}
  |\overline{K}_{\cal G}\ra=\int D\phi |\phi_1-\phi_2, \phi_2-\phi_3, \phi_4-\phi_3, \phi_4-\phi_1, \phi_4-\phi_2\ra_{a,b,c,d,e}\otimes |\phi_1, \phi_2, \phi_3, \phi_4\ra_{1,2,3,4}.
\end{equation}
The stabilizers of this extended Kitaev state are completely
different from the simple Kitaev state. In fact they are: For
each vertex $v\in V$,
\begin{equation}\label{Avtilde3}
  {C}_v:= X_v \prod_{e \in E_v} X_{e}^{\pm},
\end{equation}
where the convention for the $\pm$ signs is the same as in Kitaev
state and for each edge $e\in E$ which goes from $v_1$ to $v_2$,
\begin{equation}\label{Avtilde2}
  {D}_e:= Z_e^{-1} Z_{v_1}Z_{v_2}^{-1},
\end{equation}
where again we have suppressed the continuous arguments of these
stabilizers for ease of notation. For the example given in figure
(\ref{fig5}), some of the stabilizers are:
\begin{eqnarray}\label{}
C_1=X_1 X_a X_d^{-1}, \ \ \ C_2=X_2X_a^{-1}X_e^{-1}X_b, \ \ \
D_a=Z_1Z_a^{-1}Z_2^{-1}, \ \ \  D_b=Z_2Z_b^{-1}Z_3, \cdots
\end{eqnarray}

\subsection{Weighted Graph States} Finally we come to the definition of continuous
variable weighted graph states. Here as in the qubit case we start
with an initial product state of the form
$|\Omega\ra=|0\ra_p^{\otimes V}$. However there is an important
difference in that on each edge, instead of the simple $CZ$
operator, we can act by the $CZ(J)$ operator where the real
parameter $J$ may depend on the edge. Therefore we obtain what we
call a weighted graph state. Denoting the collection of all
weights by $J$, we have
\begin{equation}\label{}
  |\Psi_G( J)\ra=\bigotimes_{e\in E}(CZ(J_e))|\Omega\ra,
\end{equation}
where $|\Omega\ra=|0\ra_p^{\otimes V}$. The explicit form of the
state will then be given by
\begin{equation}\label{}
  |\Psi_G(J)\ra=\int D\phi e^{\sum_{<i,j>}J_{ij}\phi_i\phi_j}|\phi_1, \cdots \phi_N\ra,
\end{equation}
where $N$ is the number of vertices and $<i,j>$ denotes the edge
connecting the vertices $i$ and $j$ carrying weight $J_{ij}$.
 Note that in contrast to a
decorated edge which is denoted by $e_{ij}$ (going from $i$ to
$j$), a weighted edge is denoted by the symmetric symbol $ \la
i,j\ra $. According to (\ref{eqcz}), the stabilizers of this state will be
of the form
\begin{equation}\label{}
  K_i(t):=X_i(t)\prod_{j\in N_i}Z_j(J_{ij}t), \h \forall \ i\in V,
\end{equation}
where in the left hand side we have suppressed the dependence on
the weights for simplicity.\\

We emphasize that definition of the Kitaev states and extended
Kitaev states require the underlying graph to be decorated, while
a graph state needs only a weighted graph for its unambiguous
definition.  We are now left with an important question of
whether there is a simple relation between the above three kinds
of states or not. The answer turns out to be positive and is
explained in the next subsection.

\subsection{Relations between the above three states}\label{sec1}
Consider the extended Kitaev state corresponding to a decorated
graph $G=(V,E,\sigma)$. The explicit form of the state is shown in
(\ref{eqeG}). From (\ref{eqeG}) it is clear that if we project all the vertices in
$V$ on the zero-momentum basis, we will arrive at the Kitaev
state for the same graph. More explicitly we have
\begin{equation}\label{}
 \la \Omega|\overline{K}_{G^{\sigma}}\ra=\frac{1}{ \sqrt{ (2\pi)^{|V|} } }|K_{G^{\sigma}}\ra,
\end{equation}
where $|\Omega\ra=|0\ra_p^{\otimes V}$, and we have explicitly
indicated the decoration $\sigma$.\\
It is instructive to understand this in an alternative way, that
is by showing that measurement in the momentum basis actually
transforms the stabilizer set of the extended Kitaev state, i.e.
$S(|\overline{K}_G^{\sigma}\ra)$ to the stabilizer set of the
Kitaev state, $S(|K_G^{\sigma}\ra)$. From the stabilizer formalism
we know that measurement of a state $|\Psi\ra$ in the basis of an
operator $M$, removes all the operators which do not commute with
$M$ from the set $S(|\Psi\ra)$ and leaves us with a smaller
subset. This subset is generated by all the original generators
or their products thereof which commute with $M$. Having this in
mind, it is straightforward to see that measurement in the
momentum (the $X$ basis) leaves all the vertex stabilizers $C_v$
intact (except of course removing the vertex $X_v$ from it),
hence changing it to $A_v$ as in (\ref{eqA}). However since $X_v$
does not commute with  $Z_v$, measurements of all the vertices
remove all the generators $D_e$. The only combinations which
survive this elimination will be their product around any faces.
These are nothing but the operators $B_f$ for all faces, which
are just the right stabilizers of
$|K_{G^{\sigma}}\ra$.\\

Let us now study the relation between the extended Kitaev states
and graph states. It turns out that there is a simple relation
between the two only if the weights of the edges incident on each
vertex add up to zero, that is, if
\begin{equation}\label{}
  \sum_{j}J_{ij}=0,\h \forall i.
\end{equation}
Since $J_{ij}=J_{ji}$, this means also that $\sum_{i}J_{ij}=0$.
In such a case we can convert an extended Kitaev state to a
weighted graph state on the same graph by suitable measurements
on the edges. To this end we proceed as follows: Let us project
each edge $e_{ij}$ of the extended Kitaev state on the state $\la
\beta_2(t_{ij})|$ defined in (\ref{eqquad}).\\

Let $\la \beta_2({\bf t})|:=\prod_{e_{ij}}\la
\beta_{2}(t_{ij})|$, then we have
\begin{equation}\label{}
\la \beta_2({\bf t})|\overline{K}_G\ra=\int D\phi
e^{\sum_{e_{ij}}-it_{ij} (\phi_i-\phi_j)^2}|\phi_1, \cdots
\phi_N\ra.
\end{equation}
If we choose the parameters $t_{ij}$ of quadratures so that
$\sum_{i}t_{ij}=\sum_j t_{ij}=0$, we find
\begin{equation}\label{46}
\la \beta_2({\bf t})|\overline{K}_G\ra=\int D\phi
e^{\sum_{<i,j>}it_{ij} \phi_i\phi_j}|\phi_1, \cdots
\phi_N\ra=:|\Psi_{G}({\bf it})\ra,
\end{equation}
which is a weighted graph state with weights $it_{ij}$
assigned to each edge $\la i,j\ra.$\\

Note that from the Kitaev state for the rectangular lattice
$|K_{\square}\ra$, the Kitaev state for any other graph $|K_G\ra$
can be obtained simply by measurement of the edge modes in the
momentum or coordinate bases. In fact projecting an edge mode on
the zero-momentum state $|0\ra_p$ removes that link from the
graph while projecting it on the zero coordinate state $|0\ra_q$
merges the two endpoints of that edge. These are shown in figure
(\ref{fig6}) and are proved as follows:
\begin{figure}
  \centering
  \epsfig{file=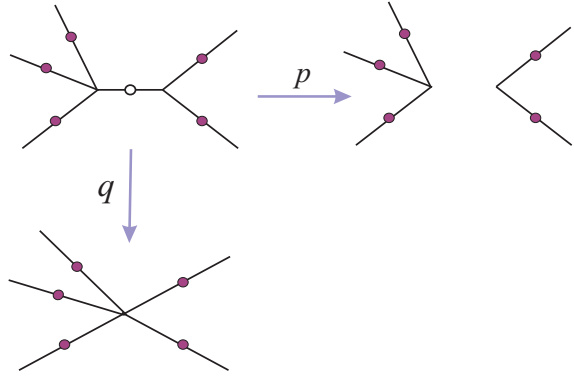,width=7.5cm}
  \caption{(Color online) The effect of measurements of an edge in an extended Kitaev state. Measuring in the momentum (X) basis
   (i.e. projecting onto the $_p\la 0|)$
   removes the edge, while measurements in the coordinate (Z) basis (i.e. projecting onto the $_p\la 0|$), merges the two end points of that edge.
    }
  \label{fig6}
\end{figure}
Let $$|\psi\ra=\int d\phi_1 d\phi_2 D'\phi |\phi_1-\phi_2\ra
\prod_{i\in L}|\phi_1-\phi_i\ra\prod_{j\in R}|\phi_2-\phi_j\ra
\cdots,$$ where $\cdots$ denotes all the edges which involve
neither the vertex $1$ nor $2$. Projecting this state on the
state $_p\la 0|$ on the edge $\la 1,2 \ra$, leaves us with
\begin{equation}\label{}
  |\psi_p\ra=\int d\phi_1 d\phi_2 D'\phi \prod_{i\in
L}|\phi_1-\phi_i\ra\prod_{j\in R}|\phi_2-\phi_j\ra \cdots,
\end{equation}
which is nothing but the same Kitaev state with the edge $\la 1,2\ra$
totally removed. On the other hand projecting $|\psi\ra$ on the
state $_q\la 0|$ on the edge $\la 1,2\ra$, leaves us with
\begin{eqnarray}\label{}
  |\psi_q\ra&=&\int d\phi_1 d\phi_2 D'\phi\delta(\phi_1-\phi_2) \prod_{i\in
L}|\phi_1-\phi_i\ra\prod_{j\in R}|\phi_2-\phi_j\ra \cdots \cr &=&
\int d\phi_1 D'\phi \prod_{i\in L}|\phi_1-\phi_i\ra\prod_{j\in
R}|\phi_1-\phi_j\ra \cdots,
\end{eqnarray}
\begin{figure}
  \centering
  \epsfig{file=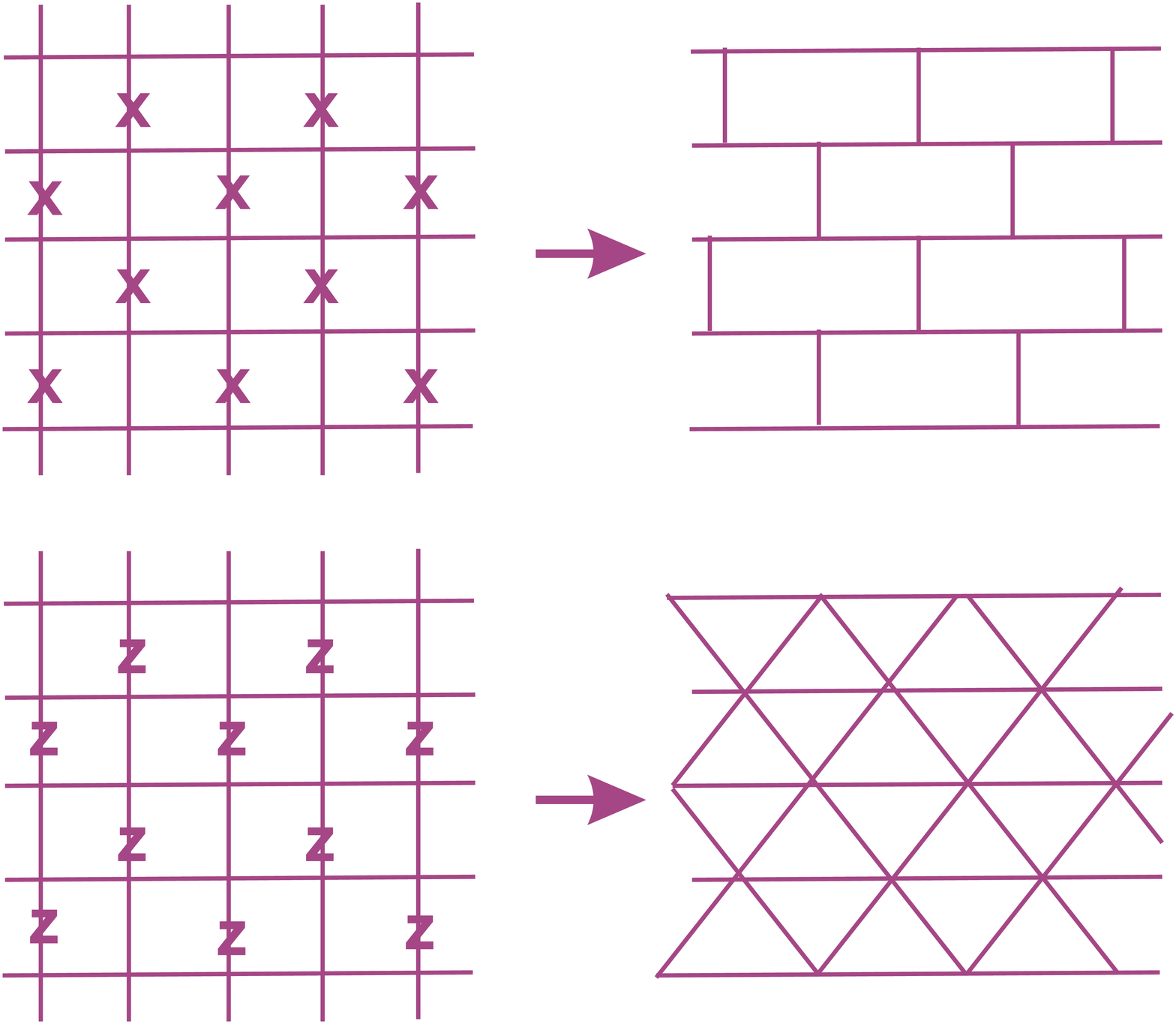,width=6.5cm}
  \caption{(Color online) The Kitaev state on the rectangular lattice, when specific edges are measured in the $X$ or $Z$ bases, will turn into the
  Kitaev state on the Hexagonal or Triangular lattices.}
  \label{fig7}
\end{figure}
which means that the two endpoints of the edge $\la 1,2\ra$ have
been merged together. With these two simple rules of deleting and
merging one can obtain the Kitaev state of any graph starting
from the one on the rectangular lattice. Figure (\ref{fig7}) shows an
important example in which measurements of some of the edges in the
momentum basis (and hence removing them), transforms
$|K_{\square}\ra$ to the Kitaev state on the hexagonal lattice
$|K_H\ra$. Measurement of the same edges in the coordinate basis
(and hence merging the two endpoints) produces a uniform lattice
whose faces are triangles, hence a triangular lattice. This is in
accord with the fact that the Hexagonal and Triangular lattice
are dual to each other, a subject which will
be explored further in the sequel.\\

Finally we use the well-known universality of cluster states
proved in the context of measurement-based quantum computation
\cite{mqc5, mqc, mqc2, mqc3, mqc6, mqc7, mqc8} to state that both the Kitaev state $|K_G\ra$ and the
extended Kitaev state $|\overline{K}_G\ra$ can be obtained by
Gaussian measurements from a sufficiently large cluster state
$|\Psi_\square\ra$. The fact that the measurements need only be
Gaussian is due to the fact that all these kinds of states are
stabilized by subgroups of $W_n$ and hence they can be converted
to each other by unitary operators belonging to the Clifford
group. Using the well-known fact from MQC, that Clifford
operators can be implemented by Gaussian measurements, we arrive
at the proof of the above statement.

\section{Quantum formalism for the partition functions}\label{3}
In this section we show how the partition function of a classical
model defined by a Hamiltonian on a continuous variables on an
arbitrary graph, can be expressed in the quantum formalism. First
consider the case where there is no local term or on-site
interaction, that is the Hamiltonian is of the form
\begin{equation}\label{eq31}
H=\sum_{\la i,j \ra}V_{ij}(\phi_{i}-\phi_{j}),
\end{equation}
where $V_{ij}(x)$ is an arbitrary function. We allow for the
function $V_{ij}$ to depend on the edge $e_{ij}$ in order to
cover also the inhomogeneous cases. The partition function of
this model is,
\begin{equation}\label{zk}
\mathcal{Z}(G,\{V\})=\int D\phi^{N}e^{-i\sum_{\la i,j \ra}V_{ij}(\phi_{i}-\phi_{j})},
\end{equation}
where $N$ is number of vertices and we have absorbed the
parameter $\beta\equiv \frac{1}{k_bT}$  in the Hamiltonian. We
will do this in all expressions of partition functions which
follow.\\

\textbf{Remark:} We have defined the partition function in the
form (\ref{zk}), in order to be able to deal with unitary
operators in the measurement based quantum computation. Dealing
with non-unitary operators does not pose any problem in the Ising
model \cite{completeness}, since states like $|\a\ra=e^{-\beta
J}|0\ra+e^{\beta J}|1\ra$ are normalizable states. In the
continuous case the analogue of the above state may be
non-normalizable, rendering the projection to such states
problematic. Instead we resort to partition functions of the type
(\ref{eqih}) with the understanding that the final results, like
dualities,  and completeness can be analytically continued to the
whole complex plane.\\

Due to the shift invariance of the Hamiltonian $\phi_i\lo
\phi_i+\xi$, the above partition function is infinite, so we have
to modify the partition function by fixing a gauge, which we will
do later on. For the present we deal with the above partition
function as it is. Defining the product state
\begin{equation}\label{eq21}
|\alpha\rangle=\bigotimes_{e_{ij}}|\alpha_{ij}\rangle ,
\end{equation}
where $$|\alpha_{ij}\rangle=\int dx e^{- iV_{ij}(x)}|x\rangle$$ is
defined on the edge $e_{ij}$, one can then write the partition
function (\ref{zk}) in the quantum formalism as
\begin{equation}\label{eqq0}
\mathcal{Z}'(G,\{V\})=\langle \alpha|K_{G}\rangle,
\end{equation}
where $|K_G\rangle$ is the Kitaev state on the graph $G$ (\ref{eqG'}). In
this way, as in the qubit case, the pattern of interactions is
encoded in the entangled Kitaev state and the strength of
interactions (including the temperature) are encoded into the
product state $\la \a|$. To fix the shift invariance we define a
gauge-fixed partition function as
\begin{equation}\label{eqq00}
\mathcal{Z}(G,\{V\})=\int
D\phi\delta(\sum_{i}\phi_i)e^{-i\sum_{\la i,j \ra}V_{ij}(\phi_{i}-\phi_{j})}.
\end{equation}
Note that other forms of gauge fixing terms are possible, but we
will deal with this simple one. Also note that the shift
invariance is also present in discrete models, however in those
cases the multiplicative factor is finite and not divergent,
hence gauge fixing is not necessary. Using the gauge-fixed Kitaev
state, we can write this in the quantum formalism as
\begin{equation}\label{zk0}
  \mathcal{Z}(G,\{V\})=\la \alpha |K^0_G\ra.
\end{equation}

Let us now consider an edge $e$ and insert the operator
$\widehat{Q}_e$ inside the inner product (\ref{zk0}). In view of (\ref{eq21}) and (\ref{zk0})
we will have
\begin{equation}\label{eqqe}
  \frac{\la \a|\Hat{Q}_e|K^0_G\ra}{\la \a|K^0_G\ra}=\la
  \phi_i-\phi_j\ra,
\end{equation}
where by $\la \ \ra$, we mean the statistical thermal average.
Similarly by acting the momentum operator on the state $\la \a|$,
we find,
\begin{equation}\label{eqpe}
  \frac{\la \a|\widehat{P}_e|K^0_G\ra}{\la \a|K^0_G\ra}=-i\la
  V'_{ij}(\phi_i-\phi_j)\ra,
\end{equation}
where by $'$ we mean derivative with respect to the argument. We
will later see an important application of these equations when
they are combined with the
topological properties of the Kitaev states. \\

Consider now the case where there are on-site interactions, then
the Hamiltonian will be
\begin{equation}\label{eq30}
H=\sum_{\la i,j \ra}V_{ij}(\phi_{i}-\phi_{j})+\sum_i W_i(\phi_i),
\end{equation}
and the above formalism will be extended as follows:
\begin{equation}\label{eqq01}
\mathcal{Z}(G,\{V\},\{W\})=\langle
\overline{\alpha}|\overline{K}_{G}\rangle,
\end{equation}
where
\begin{equation}\label{eq20}
|\overline{\alpha}\rangle=\bigotimes_{e_{ij}}|\alpha_{ij}\rangle\bigotimes_{i\in
V} |\a_i\ra,
\end{equation}
in which  $$|\a_i\ra=\int dx e^{-iW_i(x)}|x\ra$$. \\

\section{Applications of the quantum formalism}\label{4}

Let us now try to understand some of the properties of a
continuous variable statistical model through the quantum
formalism. Certainly the results that we will find, like duality,
can also be derived by other means, without resorting to the
quantum formalism, however this scheme make these properties and
their derivation much more transparent.

\subsection{Correlation functions} 
\begin{figure}
  \centering
  \epsfig{file=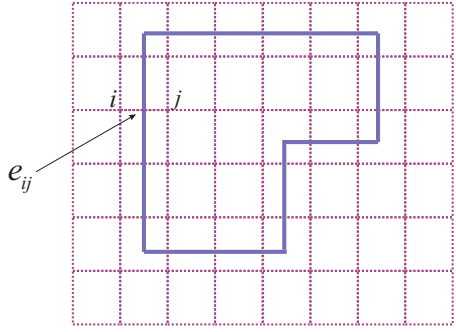,width=6cm}
  \caption{(Color online) For any closed loop, which is the boundary of a region, equation holds for the correlation functions.  }
  \label{fig8}
\end{figure}

Consider a graph $G=(V,E)$
and a Hamiltonian $H_0$ defined on it without local potentials,
$W_i(\phi)=0$.  Let $W^x_{\widetilde{C}}$ be a closed loop on the
dual graph, which is homologically trivial, (figure (\ref{fig8})) i.e. it
is the boundary of an area. Since $W^x_{\widetilde C}$ can be
written as a product of operators $A_s$, for $s$ inside the loop
$\widetilde{C}$, we have
\begin{equation}\label{}
  W^x_{\widetilde {C}}|K^0_G\ra=|K^0_G\ra,
\end{equation}
or equivalently
\begin{equation}\label{}
  \sum_{e\in {\widetilde {C}}}\widehat{P}_e|K^0_G\ra=0.
\end{equation}
Then in view of (\ref{eqpe}), this means that
\begin{equation}\label{}
  \sum_{e_{i,j}\in {\widetilde {C}}}\la V'_{ij}(\phi_i-\phi_j)\ra=0.
\end{equation}
This is a general non-trivial relation which is valid for any kind
of interaction $V$ and without using the quantum formalism, it
would have been difficult to obtain it.  Note that the other kind
of loop operator, $W^z_C$ defined as $W^z_C:=\prod_{i\in C}Z_i$
where $C$ is a loop in the graph, doesn't lead to a non-trivial
relation since in view of (\ref{eqqe}), insertion of this
operator into the inner product leads to the quantity
$\sum_{e_{ij}}(\phi_i-\phi_j)$ which identically vanishes.

\subsection{Duality} 
\begin{figure}
  \centering
  \epsfig{file=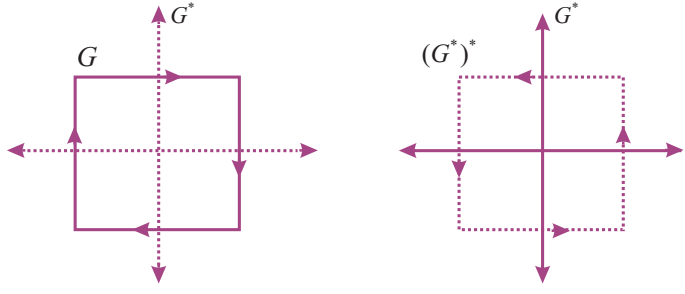,width=9cm}
  \caption{(Color online) The dual of a the dual of a graph is the same graph with all the orientations reversed. }
  \label{fig9}
\end{figure}

Denote the dual graph by $\widetilde{G}$.
The vertices, edges and faces of $G$ are in one to one
correspondence with the faces, edges and vertices of
$\widetilde{G}$ respectively. For an oriented graph we should
also choose a convention for choosing the orientations. We choose
the convention that for each edge $e$ the dual $\widetilde{e}$ be
such that the pair $(e,\widetilde{e})$ form a right handed frame,
as in figure (\ref{fig9}).  In view of the form of stabilizers of the
Kitaev states (\ref{eqA},\ref{eqB}) and the relations (\ref{Had2}), and the normalization of
the state $|K_G\ra$, we see that
\begin{equation}\label{}
|K_{\widetilde{G}}\ra=(2\pi)^{\frac{|E|}{2}}H^{\otimes
E}|K_{G}\ra.
\end{equation}
Note that contrary to the qubit case the duality relation is not
an involution, that is, as shown in figure (\ref{fig9}), the dual
of the dual of an oriented graph is not the original graph, but
the original graph with all the orientations reversed. This is in
accord with the fact that $H^2\ne I$ and indeed the action of
$H^2$ on all edges reverses their orientations. Consider now the
partition function on $G$, with $\beta H$ as in (\ref{eq31}). We
have
\begin{equation}\label{}
Z(G,\{V\})=\la \a|K_G\ra=\frac{1}{(2\pi)^{\frac{|E|}{2}}}\la \a
|{H^{\dagger}}^{\otimes
E}|K_{\widetilde{G}}\ra=\frac{1}{(2\pi)^{\frac{|E|}{2}}}\la
\widetilde{\a}|K_{\widetilde{G}}\ra,
\end{equation}
where
\begin{equation}\label{}
  |\widetilde{\a}\ra=H|\a\ra=H\int dx
  e^{-iV(x)}|x\ra=\frac{1}{\sqrt{2\pi}}\int dxdy
  e^{ixy-iV(x)}|y\ra=\int dy
  e^{-i\widetilde{V}(y)}|y\ra,
\end{equation}
where
\begin{equation}\label{eqdual}
  e^{-i\widetilde{V}(y)}=\frac{1}{\sqrt{2\pi}}\int dx e^{ixy-iV(x)}.
\end{equation}

This gives the following duality relation
\begin{equation}\label{}
Z(G,\{V\})=\frac{1}{(2\pi)^{\frac{|E|}{2}}}Z(
{\widetilde{G}},\{\widetilde{V}\}),
\end{equation}
where $e^{-i\widetilde{V}}$ is the Fourier transform of $e^{-iV}$,
as given in (\ref{eqdual}). An example of interest is when
\begin{equation}\label{}
  V_{ij}(x)=\frac{1}{2}k_{ij}x^2,
\end{equation}
which leads to the following duality relation
\begin{equation}\label{}
Z(G,\{k_{ij}\}) =
\frac{1}{(2\pi)^{\frac{|E|}{2}}}\frac{1}{\sqrt{\prod_{ij}k_{ij}}}
Z({\widetilde{G}},\{\frac{1}{k_{ij}}\}).
\end{equation}

\section{Completeness of two dimensional $\phi^4$ model for all
discrete scalar field theories}\label{5}

In this section we use the quantum formalism to show that the
discrete form of two dimensional $\phi^4$ field theory is
complete. Before proceeding we should make precise the meaning of
the above statement. By the two dimensional discrete $\phi^4$
theory, we mean the following Hamiltonian on a two dimensional
square lattice with periodic boundary conditions
\begin{equation}\label{}
  H_c=\sum_{\la r,s\ra}K_{r,s}(\phi_r-\phi_s)^2+\sum_{r}h_r \phi_r +m_r \phi_r^2+ q_r \phi_r^4,
\end{equation}
where $K_{r,s}\in \{i,-i\}, \ i=\sqrt{-1}$ and the real
parameters $h_r$, $m_r$ and $q_r$ denote respectively the
inhomogeneous external field, the quadratic (mass term) and
quartic coupling strengths. The linear terms $\{h_r\}$ are also
necessary for completeness. Denote the partition function of this
model by
\begin{equation}\label{eqih}
  Z(\sq, \{h\}, \{m\},\{q\}):=\int D\phi e^{-iH_c},
\end{equation}
where $\sq$ is a 2D square lattice of size $N$,  and $\{h\}$,
$\{m\}$ and $\{q\}$ denote the totality of all the inhomogeneous
coupling strengths. By completeness we mean that the partition
function of any other model
\begin{equation}\label{eqgener}
  H=\sum_{ r,s}V_{r,s}(\phi_r-\phi_s)+\sum_r W_r(\phi_r),
\end{equation}
on any graph $G=(V,E)$ is equal to the partition function for the
Hamiltonian $H_c$ on the 2D square lattice, for some specifically
chosen sets $\{h\}$, $\{m\}$ and $\{q\}$. \\

Note that the interaction terms in the Hamiltonian $H$ are not
necessarily of nearest neighbor type. In fact as we will show in
the proof, $H$ can even contain $k-$ body interactions, as long
as $k$ is bounded above by a finite constant independent of
$|V|$.\\

Consider a general Hamiltonian of the form (\ref{eqgener}). The partition
function of this model can be written in the quantum formalism as
\begin{equation}\label{}
  Z(G,\{V\},\{W\})=\la \overline{\a} |\overline{K}_G\ra,
\end{equation}
where
\begin{equation}\label{}
  \la \overline{ \a}|=\bigotimes_{e_{i,j}}\la \a_{ij}|\bigotimes_i \la \a_i|,
\end{equation}
and
\begin{equation}\label{}
  \la\a_{ij}|=\int dy e^{-iV_{ij}(y)}\la y|,\h \la\a_{i}|=\int dy e^{-iW_{i}(y)}\la y|.
\end{equation}
Note that $\la \a_{ij}|$ lives on the edge vertex $v(e_{ij})$ and
$\la \a_i|$ lives on the vertex $i$.  Rewriting the above states
in the form
\begin{equation}\label{}
  \la\a_{ij}|=\int dy \la y|e^{-iV_{ij}(\widehat{Q})}=\la 0_{_p}|e^{-iV_{ij}(\widehat{Q})},\h \la\a_{i}|=\int dy \la y|e^{-iW_i(\widehat{Q})}=\la 0_{_p}|e^{-iW_i(\widehat{Q})},
\end{equation}
where $\la 0_{_p}|$ is the zero momentum eignestate, we find an
equivalent form for the partition function, namely
\begin{equation}\label{}
  Z(G,\{V\},\{W\})=\la \ {\bf 0_{_p}} |\bigotimes_{e_{ij}}  e^{-iV_{ij}(\widehat{Q})}  \bigotimes_i
  e^{-iW_i(\widehat{Q})}|\overline{K}_G\ra,
\end{equation}
where $\la {\bf 0_{_p}}|=\bigotimes_{e_{ij}\in E}\la
0_p|\bigotimes_{i\in V}\la 0_{_p}|$ is the product of all
zero-momentum states on the edge vertices and ordinary vertices
of the graph $G$. We now note that according to (\ref{eqquad}),
we can approximate the unitary operators $
e^{-iV_{ij}(\widehat{Q})} $ and $e^{-iW_i(\widehat{Q})}$ to any
degree of precision by a product of the operators $H$ (the
Hadamard) , $e^{-it\widehat{Q}}$, $e^{-it\widehat{Q}^2}$ and
$e^{-it\widehat{Q}^4}$. As explained in (\ref{sec0}),
implementation of these operators on a state is effected by
suitable possibly non-Gaussian measurements (i.e. projections of
vertex modes on the states $\la \beta_i(t)|( i=1, 2, 4)$) of an
appropriate enlargement of that state, i.e. one simply adds
necessary nodes, glues them by $CZ$ operators and measures the
additional nodes as exemplified in figure (\ref{fig2}) to affect
a desired unitary gate on the original qumode of the lattice. Let
us denote this intermediate graph by $G'$, its associated state
by $|\Psi_{G'}\ra$, and the collection of all necessary
measurements on it by $\la \beta_{1,2,4}|$, then we will have
\begin{equation}\label{78}
\bigotimes_{e_{ij}}  e^{-iV_{ij}(\widehat{Q})}  \bigotimes_i
  e^{-iW_i(\widehat{Q})}|\overline{K}_G\ra=\la \beta_{1,2,4}|\Psi_{G'}\ra,
\end{equation}
The configuration of the graph $G'$ may be complicated, but the
important point is that $|\Psi_{G'}\ra$ is nothing but a
stabilizer state and hence in principle it can be obtained from a
cluster state by Gaussian measurements $\la \b_i(t)| (i=1,2)$.
Therefore we have
\begin{equation}\label{79}
|\Psi_{G'}\ra=\la \beta_{1,2}|\Psi_\sq\ra.
\end{equation}
Note that up to now all the projections have been performed on
the vertices of the cluster state. The cluster state
$|\Psi_\square\ra$ can be a weighted cluster state where the
weights of all edges are $\pm 1$ and for each vertex the weights
of all edges add up to $0$. Such a cluster state can be obtained
from an extended Kitaev state on the rectangular lattice by
projecting all the edge-vertices on the states $\la \beta_{2}(\pm
1)|=\int dy e^{\mp i y^2}\la y|$ according to whether the weights
of the edges are $+1$ or $-1$.

This is the only place where projections are made on the edge
vertices, and in fixed directions (i.e. eigenstates of $ XZ^{\pm
1}$). We write this symbolically in the form
\begin{equation}\label{80}
|\Psi_{\square}\ra=\la \pm|\overline{K}_\square\ra.
\end{equation}
Combining (\ref{78}), (\ref{79}) and (\ref{80}) we finally arrive
at
\begin{equation}\label{}
  Z(G,\{V\},\{W\})=\la {\bf 0_{_p}}|\la \beta_{1,2,4}, \pm_{e} |\overline{K}_\square\ra,
  \end{equation}
where $\pm_{e}$ encapsulates all the measurements $XZ^{\pm}$
which are performed on the edges of the extended Kitaev state,
and $\la \beta_{1,2,4}|$ represents all the projections $\la
\beta_i|$ for $i=1,2,4$ on the vertices.  Putting all this
together we finally arrive at the result that
\begin{equation}\label{}
  Z(G,\{V\},\{W\})=Z(\sq, \{h\},
\{m\},\{q\}).
  \end{equation}
It is important to note that since the edge vertices are measured
in the $ZX^{\pm}$ bases and the resulting edge is projected on
the eigenstates $|\pm\ra |\propto \int e^{\pm i y^2}|y\ra dy$,
the interactions between neighboring vertices in the complete
model is restricted to be of the type $\pm i(\phi_i-\phi_j)^2$.
In this way all the couplings in the original model have been
transferred to the mass and potential terms on the vertices.\\
 As a byproduct this argument shows that when the original model has only quadratic couplings (i.e. it is free field), then there is no linear or quartic coupling in  the  model on the rectangular lattice which reduces to this model by Gaussian measurements. This means that the free field theory on the 2D rectangular lattice is complete and can produce any other field theory on any lattice. \\
It is a simple matter to show that the $\phi^4$ theory can also
reproduce models with $k-$ body interactions. We know from
\cite{u1}, that the 4D $U(1)$ lattice gauge theory is complete.
Therefore it is enough to show that $\phi^4$ theory can reduce to
$4D$ $U(1)$ lattice gauge theory. The Hamiltonian of the latter
model is given by

\begin{equation}\label{u1H}
H=-\sum_{p}J_p\cos (\phi_1-\phi_2-\phi_3+\phi_4),
\end{equation}
where $p$ denotes a plaquette, $J_p$ denotes the coupling
constant on $p$, and $\phi_i$'s are the continuous variables
around $p$. The indices $1,2,3$ and $4$ denotes the edges of $p$
when traversed in clockwise direction. The point is that the
partition function of such a model can again be written as a
scalar product $Z=\la \a|G\ra=$, with $\la \a|$ a product state
over all plaquettes,
\begin{equation}\label{}
  \la \a|=\otimes_{p}\la \a|_p, \h \la \a|_p:=\int dy e^{-iJ_p\cos
  y}\la y|
\end{equation}
and $|G\ra$ a new stabilizer state,
\begin{equation}\label{}
  |G\ra=\int Dx |\cdots, (x_1-x_2-x_3+x_4)_p, \cdots\ra.
\end{equation}
If we now note that $\la \a|_p$ can be written as $ \la \a|_p=\la
0|_p e^{-iJ_p\cos
  \hat{Q}}$ and the latter operator can indeed be expanded to any
  desired accuracy in terms of $e^{-it\hat{Q}^2}$ and
  $e^{-it\hat{Q}^4}$, the assertion will be proved along the same
  line as indicated above.\\

{\it Efficiency:} It is shown in \cite{quantum formalism} that
simulating the partition function of any Ising or Potts type
model on an arbitrary graph can be done on the Ising model on a
square lattice with only a polynomial overhead in the number of
spins. For more general models however, an exponential overhead
may be needed. A similar statement is true also in our case,
namely for simulating the partition function of a model with
nearest neighbor interactions on a graph with $N$ vertices, we
need a $\phi^4$ model on a cluster state with $P(N)$ vertices
where $P(N)$ is a polynomial of $N$. This result is a combination
of the universality result of cluster states which by only a
polynomial overhead can produce any other quantum state and the
fact that any quantum unitary $e^{iV(\hat{q})}$ can be decomposed
to a product of polynomial number of unitaries of the form
$e^{it\hat{q^n}}$, with $n=1,2,4$ to any degree of precision.
This result is also true when each site interacts with a finite
number $k$ of its neighbors, where $k$ is independent of $N$.
For more general models, an exponential overhead in the number of sites, will be necessary like the case of Ising model.  \\

\section{Discussion} The concept of completeness of certain
statistical models is very fascinating. The idea that in
principle a single complete model, like the 2D Ising model in its
rich phase structure, various phases of all the other models,
regardless of their lattice structure, type of order statistical
variables and order parameter, and the interactions, is an idea
which needs much exploration in the future. One of the basic
questions is that what other types of models are complete. In
this paper we have shown that the $\phi^4$ field theory is a
complete model. Like the case of Ising model \cite{completeness}
or the $U(1)$ lattice gauge theory \cite{u1}, our proof is an
existence proof at present. The next step in such a program will
be to show how other specific models can be obtained from these
complete models and what insight about them can be obtained. Like
the existence proof itself, this step relies also heavily on
techniques from quantum information theory, notably the
measurement-based quantum computation. In particular it will be
very desirable to formulate an algorithmic approach for deriving
any specific model (its graph structure and coupling strengths)
from a complete model. In this way apparently unrelated models
will be linked to each other and the insight gained from this
approach will have far reaching consequences in statistical
mechanics, exactly solvable models and critical phenomena.\\

\textbf{Acknowledgements:} The authors would like to thank A.
Rezakhani, L. Memarzadeh for valuable discussions. V. K. would
also like to thank Abdus Salam ICTP for partial support through
the regular associate award.

\end{document}